# Distributed Asymmetric Allocation: A Topic Model for Large Imbalanced Corpora in Social Sciences

Kohei Watanabe (Waseda University)

August 2025

## Abstract

Social scientists employ latent Dirichlet allocation (LDA) to find highly specific topics in large corpora, but they often struggle in this task because (1) LDA, in general, takes a significant amount of time to fit on large corpora; (2) unsupervised LDA fragments topics into sub-topics in short documents; (3) semi-supervised LDA fails to identify specific topics defined using seed words. To solve these problems, I have developed a new topic model called distributed asymmetric allocation (DAA) that integrates multiple algorithms for efficiently identifying sentences about important topics in large corpora. I evaluate the ability of DAA to identify politically important topics by fitting it to the transcripts of speeches at the United Nations General Assembly between 1991 and 2017. The results show that DAA can classify sentences significantly more accurately and quickly than LDA thanks to the new algorithms. More generally, the results demonstrate that it is important for social scientists to optimize Dirichlet priors of LDA to perform content analysis accurately.

## Keywords

**Introduction**

Social scientists have long been analyzing topics or themes of documents to understand important issues. As large textual data became more accessible, thanks to online news, social media, and digital archives, many found topic models very useful because their unsupervised algorithms allow users to identify topics in large corpora without much human involvement. Among various topic models, latent Dirichlet allocation (LDA) (Blei et al. 2003) has been arguably the most popular topic model in social sciences.

Unlike latent semantic analysis (LSA) (Deerwester et al. 1990) and probabilistic LSA (Hofmann 1999), LDA involves modeling the data-generating process, in which an author chooses words to create a desired mix of topics in documents. It assigns topics to individual words through iterative sampling based on co-occurrences of words in documents. More recent additions to topic models are Top2vec (Angelov 2020) and BERTopic (Grootendorst 2022). These algorithms classify documents in a lower dimensional space created using word embedding techniques. These new topic models are becoming increasingly popular among social scientists. In fact, recent studies have shown that they are more capable than LDA in identifying topics (Egger and Yu 2022; Gan et al. 2024), but it is too early to dismiss LDA as obsolete. LDA is transparent and independent since its algorithm does not rely on pre-trained word vectors. It is also flexible because it permits extensions to solve different types of problems. The original algorithm was modified to address the sparsity of word co-occurrences in short documents (Amoualian et al. 2016; Du et al. 2012; Gruber et al. 2007; Jiang et al. 2019; Yan et al. 2013; Watanabe and Baturo 2023), the fragmentation or agglomeration of topics (Chien et al. 2018; Syed and Spruit 2018; Wallach et al. 2009), and the high computational costs of iterative sampling (Newman et al. 2009; Smyth et al. 2008; Nutakki et al. 2014).

These proposed algorithms enable LDA to become much more useful for social scientists. However, the lack of implementations in accessible software packages has led to the widespread use of the original algorithm without experiencing the advanced capabilities of the proposed alternatives.[1] Therefore, I have developed an enhanced LDA called *distributed asymmetric allocation* (DAA) that combines algorithms for distributed computing (Newman et al. 2009), semi-supervised learning (B. Lu et al. 2011), sequential sampling (Watanabe and Baturo 2023), Dirichlet prior adjustment, and convergence detection, all of which are implemented in an open-source software package in this study.[2]

Further, earlier discussions on topic models often centered around the optimization of the number of topics (Arun et al. 2010; Cao et al. 2009; Deveaud et al. 2014; Griffiths and Steyvers 2004; Watanabe and Baturo 2023). This led many users of LDA to focus only on the optimization of the number of topics, ignoring its other parameters. Even if they are aware of the importance of asymmetric Dirichlet priors to identify topics accurately, the lack of automated methods for optimization in widely available tools prevent them from doing so. This study aims to address these problems through the publication of the open-source software package.In the following sections, I first identify common problems that social scientists face when they employ LDA in their research. Second, I explain the algorithms of DAA that can solve these problems. Third, I apply DAA and LDA to classify sentences from the United Nations General Assembly speech corpus (Baturo et al. 2017) to evaluate their impact. Fourth, I demonstrate how DAA can lead to more plausible conclusions in content analysis thanks to its more accurate estimation of topic frequencies.

---

[1] For example, the *Gensim* package for Python does not support optimization of asymmetric priors when distributed computing is enabled. The *topicmodels* package for R neither performs distributed computing nor estimates asymmetric priors.
[2] The data and scripts for this study are made available at https://doi.org/10.7910/DVN/***.

The results of the evaluation show that DAA can classify sentences more accurately and quickly than LDA: the F1 scores are higher by 0.21 points in DAA when both Dirichlet prior adjustment and sequential sampling are enabled. Simultaneously, the execution time is roughly 20 times shorter in DAA when both distributed computing and convergence detection are used. These improvements will enable social scientists to perform topic analysis of large imbalanced corpora without waiting long hours.

The results of the content analysis reveal substantive differences between DAA and LDA. When they are applied to classify the sentences of the speeches at the United Nations General Assembly, the frequencies of topics in DAA vary more widely during the post-Cold War period. Moreover, these changes more strongly correspond to the occurrences of key political events thanks to the asymmetric Dirichlet priors.

**Problems**

Social scientists often find it difficult to use LDA in their research because (1) it takes a significant amount time to identify topics in a large corpus (Newman et al. 2009); (2) it fragments topics into sub-topics in a corpus of short documents (Nutakki et al. 2014; Lin 2023); (3) it fails to identify highly specific topics defined using seed words (Watanabe and Baturo 2023).

LDA takes a considerable amount of time to identify topics in a large corpus because it employs an iterative algorithm and collapsed Gibbs sampling to assign the most likely topic for each word. Furthermore, its computational cost grows proportionally to the total number of words in the corpus, the number of topics to identify, and the number of iterations (Heinrich 2008). This often leads to a quadratic increase in the execution time because the diverse content of a large corpus requires many topics to be identified. The execution time can be shortened by Gibbs sampling on multiple processors (Newman et al. 2009), but distributed computing is not very effective when the number of topics is small.

LDA tends to fragment topics into sub-topics in a corpus of short documents because it cannot accurately infer the overall probability of topics through the sampling of topics in individual documents. Usually, the frequencies of topics in long documents correlate with their overall frequencies, but they do not in short documents (e.g., sentences and social media posts) because short documents only contain words for a few related topics (Yan et al. 2013). The use of asymmetric Dirichlet priors helps LDA to classify short documents, but it is difficult for the users to manually set the hyper-parameters.

Semi-supervised LDA fails to identify highly specific topics defined using seed words because seed words do not offer information on the frequencies of topics. Seed words increase the chance that their co-occurring words receive desired topics by inducing bias through pseudo-counts in topic assignments (B. Lu et al. 2011). The use of unequal numbers of seed words (or their matches) for topics informs the Gibbs sampler about their frequencies, but the number of seed words is usually determined by the complexity or broadness of the topics instead of their frequencies (Watanabe and Zhou 2020).

**Algorithms**

LDA infers parameters from the distributions of documents and words through Gibbs sampling. In Figure 1, the most important variables are topics, $Z_h$, and words used to express these topics, $W_h$. The Gibbs sampler iteratively assigns the most likely topics to each of the $Z_h$ based on the $\theta$ and $\phi$ distributions, whose shapes are determined by the Dirichlet priors, $\alpha_k$ and $\beta_k$, respectively. The values of the priors are equal for all the topics, $k = \{1, 2, \cdots, K\}$, in symmetric models but unequal in asymmetric models.

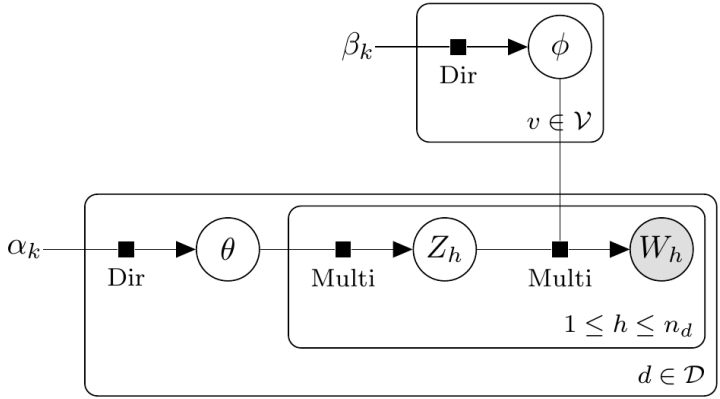

**Figure 1.** Graphical model of a simple LDA. Gray circles are variables whose values are known, whereas white circles are latent variables whose values are unknown. $\theta$ and $\phi$ are the $|D| \times K$ and $K \times |V|$ matrices, respectively; $Z$ is a vector to record topics words, $W$, in document $d$; $d$ is one of the documents ($d \in D$); $v$ is one of the unique words ($v \in V$) in the corpus.

To infer the document-topic distribution $\theta$ and the topic-word distribution $\phi$, assigned topics are saved in $M_{dk}$ and $N_{kv}$. The former is the frequency of topic $k$ found in document $d$, the latter is the frequency of topic $k$ assigned to unique word $v$, and $\alpha_k$ and $\beta_k$ are the Dirichlet priors, which are added to the frequency counts to smooth the distributions in $\theta$ and $\phi$.[3] The Gibbs sampler assigns topics to the words in the corpus based on the sampling distribution, $G$, derived as a product of $\theta$ and $\phi$. These relationships between are defined as follows:

$$G = P(Z = k|W = v, d) \propto \theta_{dk} \phi_{kv}$$

$$\theta_{dk} = P(Z = k|d, \alpha_k)$$

---

[3] For example, if $\alpha_k = 0.5$, it is assumed that topic $k$ appears at least 0.5 times in any document; if $\beta_k = 0.1$, it is assumed that all the words appear at least 0.1 times in topic $k$.

$$= \frac{M_{dk} + \alpha_k}{M_{d.} + \sum \alpha_k}$$

$$\phi_{kv} = P(W = v | k, \beta_k)$$

$$= \frac{N_{kv} + \beta_k}{N_{k.} + \sum \beta_k}$$

In Figure 2, $z_i$ is the topic assigned for word $h$ in document $d$ by the Gibbs sampler at iteration $i$. A simple LDA completes the inference by repeating sampling of topics until $i$ reaches a fixed number of iterations, $max\_iter$, which is usually between 1000 and 3000.

---

**initialize**
  randomly assign topics to $Z$
**for** $(1 \leq i \leq max\_iter)$ {
  sample topic for words: $z_i \leftarrow G(N_{kv}, M_{dk}, \alpha_k, \beta_k)$
}

---

**Figure 2.** Pseudo-code for a simple LDA. Gibbs sampling is repeated $max\_iter$ times.

DAA is created by extending the simple LDA. I first implemented distributed LDA because its parallel Gibbs sampling determines how other algorithms function. Next, I added seeded LDA, sequential LDA, the Dirichlet prior optimization, and the convergence detection algorithms to the model. Since all these algorithms are an extension of LDA, they can be enabled or disabled independently from each other. The model is seeded DAA if all the extending algorithms are enabled, but it becomes a plain-vanilla LDA if all of them are disabled. Therefore, the main challenge in developing DAA is significantly speeding up LDA while generalizing its original algorithm for extra capabilities.

The first algorithm was initially developed for approximately distributed LDA to increase the ability of LDA to process very large corpora (Newman et al. 2009). The second and third were for seeded sequential LDA to weakly supervise LDA using seed words in the classification of sentences (Watanabe and Baturo 2023). The fourth is a low-cost algorithm for automatically adjusting Dirichlet priors to avoid fragmentation or agglomeration of topics; the fifth is a simple algorithm for convergence detection to minimize the computational costs of LDA. While the first three algorithms are adopted from earlier studies, the last two are newly developed in this study.

Following the notations in Watanabe and Baturo (2023), I concisely explain the algorithms with the aid of pseudo-code (Figures 2–5). DAA is implemented in C++ using the Intel Thread Building Blocks library and published as part of the *seededlda* package for R (available on CRAN).

**Distributed Computing**

LDA is computationally expensive because its algorithm assigns the most likely topic, $k$, to each word in the corpus based on the sampling distribution derived from $\theta$ and $\phi$. Therefore, an analysis of a larger corpus not only increases the frequency of sampling proportionally to the total number of words in the corpus, $N$, but also the number of parameters proportionally to the number of topics, $K$, the number of documents, $D$, and the size of the vocabulary, $V$.

Among several distributed LDA algorithms, I chose the approximately distributed LDA (Newman et al. 2009) for its simplicity and generalizability. It assigns topics to words in the same way as the simple LDA, but it splits data into smaller chunks and performs Gibbs sampling on multiple processors in parallel (Figure 3). While $N_{kv}$ and $M_{dk}$ are global variables shared by all the processors, each subprocess $e = \{1, 2, \cdots E\}$ has a local variable to record the topic assignment $\acute{N}_{kv}^e$. In every 10 iterations, these local counts are added to the global variable, $N_{kv}$, to synchronize the topic assignment between the processors.

```
initialize
      randomly assign topics to Z
for (1 ≤ i ≤ max_iter/10 ) {
      assign D × batch_size documents to processor e
      parallel_for (1 ≤ j ≤ 10) {
          sample topic for words in the batch: z_j ← G(N_kv, M_dk, α_k, β_k)
          return Ń_kv^e
      }
      synchronize topic-word count:  N_kv ← N_kv + Ń_kv^1 + Ń_kv^2 + ⋯ Ń_kv^e
}
```

**Figure 3.** Pseudo-code for the enhanced LDA with distributed computing. Gibbs sampling is repeated $max\_iter$ times in separate processors over $D \times batch\_size$ documents; $D$ is the total number of documents in the corpus.

**Dirichlet Prior Optimization**

LDA is often fitted with symmetric Dirichlet priors, $\alpha$ and $\beta$, because the users lack the knowledge of topic distributions. Although symmetric priors are usually inappropriate (Chien et al. 2018), the algorithm can still identify topics when the conditional and the marginal probabilities of topics are similar, $P(Z = k|d) \sim \mathrm{P}(Z = k)$. This is often the case in a corpus of long documents, where topics can occur multiple times, but this is rarely the case in a corpus of short documents, where topics can occur only a few times.

When the condition is unsatisfied, the Gibbs sampler requires asymmetric priors for $\theta$, $\{\alpha_1, \alpha_2, \cdots, \alpha_k\}$, to accurately assign topics.[4] Since it is very difficult for the users to specify the parameters manually, I propose a new low-cost algorithm to set asymmetric priors, $\{\hat{\alpha}_1, \hat{\alpha}_2, \cdots, \hat{\alpha}_k\}$, by automatically adjusting the initial value, $\alpha$, for each topic. This allows the algorithm to update the priors for iteration $i$ based on the posterior probability of topics in iteration $i - 1$ (Figure 4).

---

[4] Following the recommendation by Wallach et al. (2009), the Dirichlet prior for $\phi$ and $\beta_k$ are left symmetric in DAA.

To limit the amount of the adjustment, the algorithm computes a small constant, $\varepsilon_k = \upsilon \frac{\alpha}{M_{.k}}$, with a hyper-parameter, $0 \leq u < 1$, based on the initial random assignment of topics to words.. If $\upsilon = 0$, the priors cannot change; if $\upsilon = 0.9$, $\hat{\alpha}_k$ can drop to 10% of the initial value. Each time the Gibbs sampler updates the assigned topics, $\varepsilon_k$ is added to the new topic and removed from the old topic to make asymmetric priors. Since a decrease (or increase) in a prior for a topic is offset by an increase (or decrease) in priors for other topics, the total sum of the asymmetric priors remains constant.

```
initialize
      randomly assign topics to Z
      compute constant for Dirichlet adjustment: εₖ ← adjust_alpha × α/M.ₖ
for (1 ≤ i ≤ max_iter/10 ) {
      Assign D × batch_size documents to processor e
      parallel_for (1 ≤ j ≤ 10) {
          sample topic for words in the batch: zⱼ ← G(Nₖᵥ, Mₔₖ, αₖ, βₖ)
          return Ńₖᵥᵉ
      }
      synchronize topic-word count:  Nₖᵥ ← Nₖᵥ + Ńₖᵥ¹ + Ńₖᵥ² + ⋯ Ńₖᵥᵉ
      adjust Dirchlet prior: αₖ ← αₖ + εₖ(Ńₖ.¹ + Ńₖ.² + ⋯ Ńₖ.ᵉ)
}
```

**Figure 4.** Pseudo-code for the enhanced LDA with parallel computing and Dirichlet prior adjustment. Adjusted $\alpha_k$ serves as Dirichlet priors in the next iteration, $i + 1$.

## Convergence Detection

In LDA, Gibbs sampling is usually repeated between 1,000 and 3,000 times because the users cannot easily determine the necessary number of iterations before fitting models. Although distributed computing helps to complete large numbers of iterations quickly, the execution time can become much shorter by only interrupting the iterations earlier when the topic assignment is stabilized.

It is difficult to detect the convergence in Gibbs sampling in general (Gelman and Rubin 1992), but the use of topic divergence or perplexity is infeasible because of the high computational costs. Therefore, I devised the *delta statistic* as a simple convergence detection criterion for DAA (Figure 5). It measures the stability of the topic assignment by comparing current and previous topics, $\delta_i = \left| z_j \neq z_{j-1} \right|$, and continues Gibbs sampling as long as the statistic is decreasing, $\delta_i \leq \delta_{i-1}$.[5] Since the statistic tends to fall quickly in the first few hundred iterations, it can reduce the computational cost dramatically. In distributed computing, the local variable, $\acute{\delta}^e = \left| z_j \neq z_{j-1} \right|$, can be obtained in the last sub-iteration, $j = 10$, and added to the global variable, $\delta_i$, to detect convergence.

---

[5] After convergence, $\delta$ tends to fluctuate around zero because words that only have weak association with others receive different topics each time.

```
initialize
      randomly assign topics to Z
      compute constant for Dirichlet adjustment: ε_k ← adjust_alpha × α/M_.k
for (1 ≤ i ≤ max_iter/10 ) {
      assign D × batch_size documents to processor e
      parallel_for (1 ≤ j ≤ 10) {
            sample topic for words in the batch: z_j ← G(N_kv, M_dk, α_k, β_k)
            if (j = 10) {
                  count topic changes: ́δ^e ← |z_j ≠ z_{j−1}|
            }
            return ́N^e_kv, ́δ^e
      }
      synchronize topic-word count: N_kv ← N_kv + ́N^1_kv + ́N^2_kv + ··· ́N^e_kv
      adjust Dirchlet prior: α_k ← α_k + ε_k(́N^1_k. + ́N^2_k. + ··· ́N^e_k.)
      aggregate topic changes: δ_i ← ́δ^1 + ́δ^2 ··· ́δ^e
      if (δ_i ≤ δ_{i−1}) {
            exit
      }
}
```

**Figure 5.** Pseudo-code for DAA. It enhances LDA by adding parallel computing, Dirichlet prior adjustment, and convergence detection. Convergence is checked after each sub-iteration by comparing the value of $\delta_i$ and $\delta_{i−1}$.

## Evaluation

I evaluated the proposed algorithms using a transcript of speeches at the United Nations General Assembly (Baturo et al. 2017), which scholars of international relations have analyzed to understand important political issues (Gurciullo and Mikhaylov 2017; Schoenfeld et al. 2018; Kentikelenis and Voeten 2021). Following earlier studies, I selected sentences of speeches (n = 444,206) delivered by delegates from 198 countries between 1991 and 2017 from the corpus and separated them into a training set (n = 441,557) and a test set (n = 2,649). I preprocess the corpus following the standard steps but without stemming to minimize feature engineering.[6] The sentences in the test set are given topic labels: "Greeting," "UN," "Security," "Human rights," "Democracy," and "Development," for which a list of seed words is prepared (Table 1).[7]

This corpus is suitable for evaluating the algorithms because it is large, sparse, and imbalanced. The large number of sentences requires distributed computing and convergence detection; the small number of tokens in sentences (on average 12.8) prevents the Gibbs sampler from inferring the overall frequency of topics in individual documents; the mix of frequent and infrequent topics demands the use of asymmetric priors to classify them accurately.

In the evaluations, I trained and tested the DAA model with different sets of hyper-parameters and measured their performance in terms of execution time, classification accuracy (F1), or goodness-of-fit (perplexity). I fitted the model to the sentences in the training set simply with different values of $k = \{5, 10, 25, 50\}$ to measure the execution time, but I did so with six seeded topics plus two unseeded topics to measure the classification accuracy or the goodness-of-fit.[8] Using the fitted models, I classified the sentences in the test set into one of the six topics and computed the F1 and perplexity scores.[9]

---

[6] I prepared the data using the *quanteda* package (Benoit et al. 2018) following the standard procedure: (1) segment speeches in the corpus into sentences; (2) tokenize the sentences; (3) remove punctuation marks, numbers, and gramatical words; (4) compound multi-word expressions in the seed words; (5) form a document-feature matrix; (6) remove features that occur less than 10 times in the entire corpus.

[7] The frequencies of topic labels are "Greeting": 5.7%; "UN": 18.3%; "Security": 43.3%; "Human rights": 6.8%; "Democracy": 3.6%; "Development": 28.0%. Seed words are adopted from Watanabe and Zhou (2020).

[8] I used Ubuntu 22.04 on Microsoft Azure Virtual Machine (Microsoft Azure Standard D16as v4) to fit the models, setting the batch size to 1% for distributed computing.

[9] Using the fitted DAA model, Gibbs sampling is performed over 100 iterations to classify the unseen documents in the test set.

| Topic | Seed words |
|---|---|
| Greeting | greet*, thank*, congratulat*, sir, express*, great*, mr, wish*, hop*, contribut*, anniversar*, welcom* |
| UN | united nations, international court*, security council, general assembly, organization*, reform*, secretary-general, resolution*, permanent member*, charter*, session*, conference* |
| Security | secur*, kill*, attack*, dispute*, victim*, peac*, terror*, weapon*, nuclear*, conflict*, war*, disarmament*, threat*, cris*, solution*, settlement*, force*, destruction*, militar*, violence*, arm*, fight* |
| Human rights | human rights, violat*, race*, dignit*, protect*, citizen*, educat*, humanitarian, child*, women, refugee*, communit*, people, respect*, responsib*, food*, health* |
| Democracy | democra*, autocra*, dictator*, vote*, represent*, elect*, leader*, president*, party, institution*, government*, law*, republic*, free*, leadership*, legal* |
| Development | develop*, market*, investment*, econom*, climate change, assistance*, sustain*, povert*, trade*, grow*, social*, environment*, prosperit*, progress*, financ*, cooperation* |

**Table 1.** Topic seed words. They combines the knowledge-based and frequency-based seed words proposed by Watanabe and Zhou (2020).

## Distributed Computing

The execution time of DAA became shorter and inversely proportional to the number of processors used in the large models, but it only changed a little in the small models. It took 46 minutes to fit the models with 50 topics when only one processor was used, but it became 8 minutes when eight processors were used. Similarly, it took 25 minutes to fit the models with 25 topics using one processor, but it became 7 minutes when eight processors were used. Importantly, the execution time is equal between the symmetric and asymmetric models, suggesting that the computational costs of Dirichlet prior optimization are extremely small.

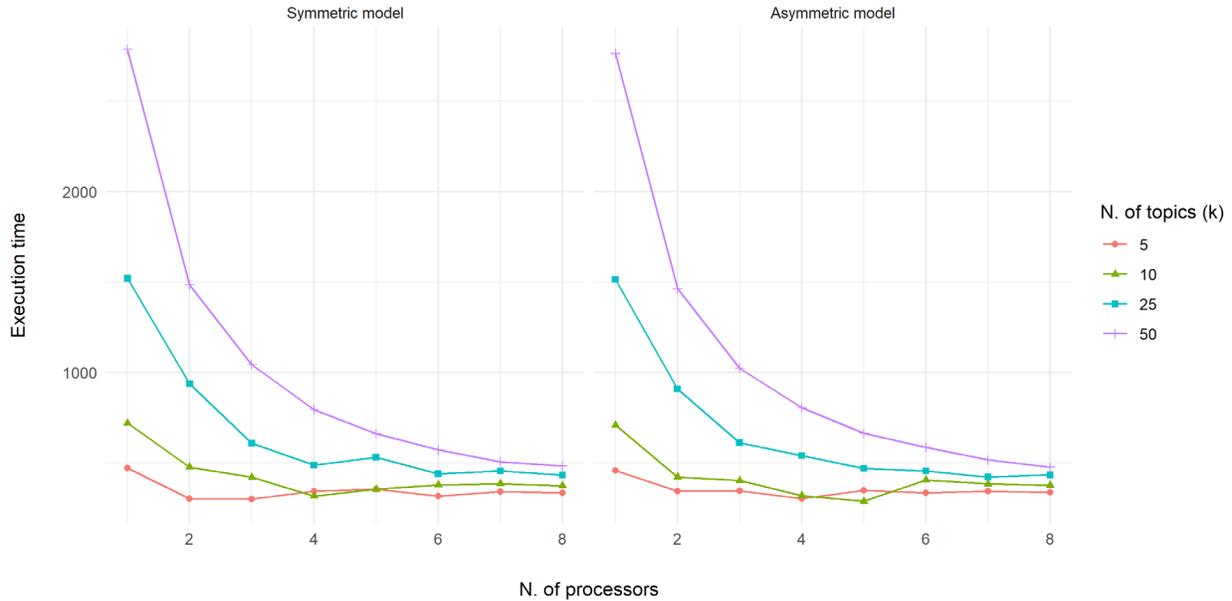

**Figure 6**. Execution time of DAA by numbers of processors. The *x*-axis is the number of processors used to fit the model; the *y*-axis is the execution time in seconds. The plotted lines represent the average values of the five models fitted under the same condition. The other hyper-parameters are $\upsilon = 0$ and $\gamma = 0.25$ for the symmetric model and $\upsilon = 0.3$ and $\gamma = 0.25$ for the asymmetric model.

**Dirichlet Prior Optimization**

The classification accuracy of DAA increased when sequential sampling was performed, but it further improved when Dirichlet priors were adjusted (Figure 7). The overall F1 score increased from 0.52 to 0.60 points when $\upsilon = 0.6$ for the weak sequential sampling ($\gamma = 0.25$); it increased from 0.57 to 0.68 points when $\upsilon = 0.3$ for the strong sequential sampling ($\gamma = 0.5$). The F1 score increased in all the topics, but the improvement was more pronounced in "Security" (+0.15), "Greeting" (+0.19), and "Democracy" (+0.09) for the strong sequential sampling. The F1 scores usually peaked when the adjustment was $\upsilon < 0.7$ for the weak sequential sampling and $0.25 \leq \upsilon \leq 0.3$ for the strong sequential sampling. The scores fell sharply if a greater adjustment was made. Interestingly, the Dirichlet prior adjustment had little or no impact on the classification accuracy of DAA in non-sequential sampling ($\gamma = 0$).

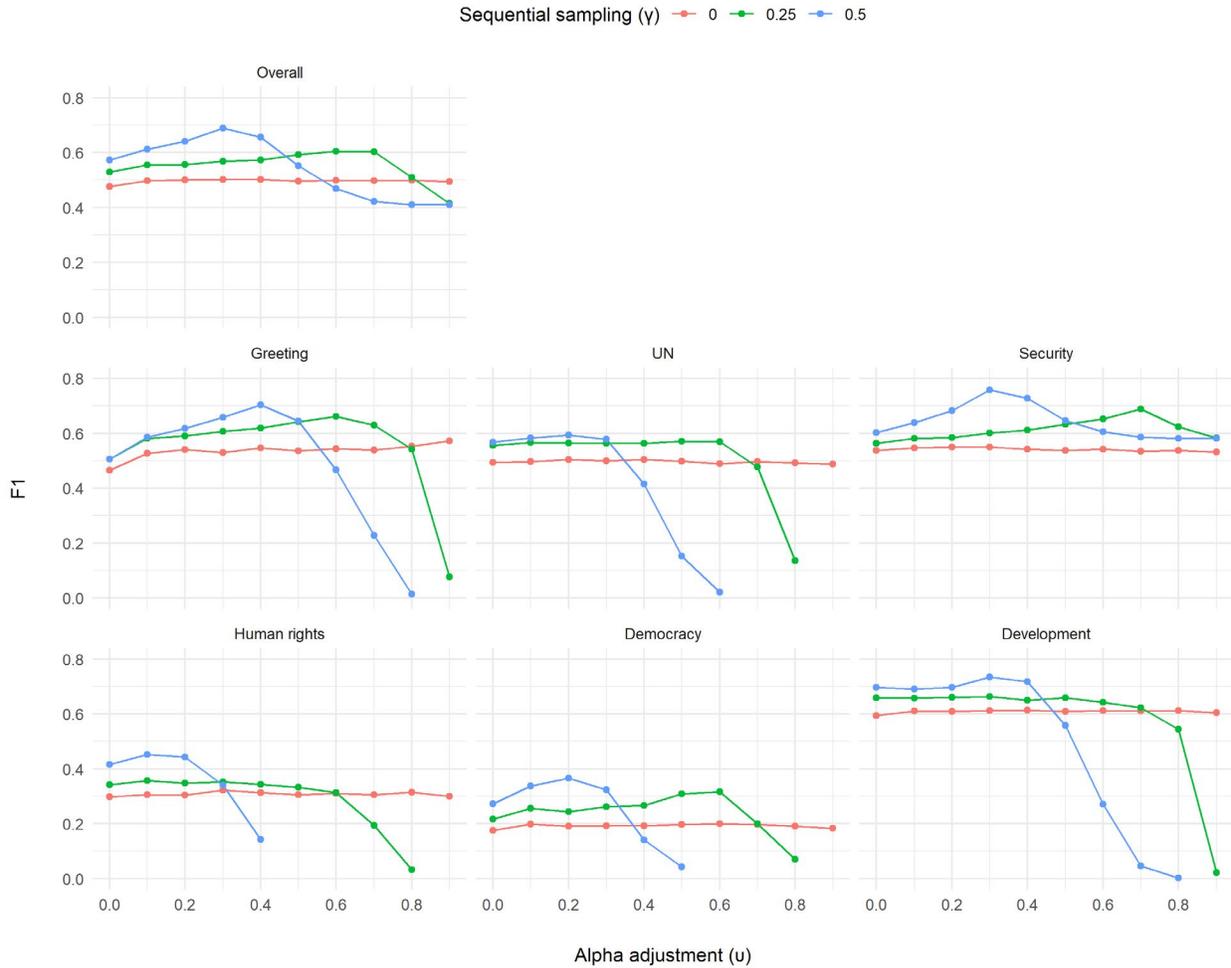

**Figure 7.** Classification accuracy of DAA by different levels of Dirichlet prior adjustment. The *x*-axis is the maximum amount of adjustment made to $\alpha_k$; the *y*-axis is the micro-average F1 score. The plotted lines represent the average values of the five models fitted under the same condition.

Dirichlet alpha adjustment improved the F1 scores dramatically because the asymmetric priors changed the chance that the topics were assigned to words (Figure 8). In strong sequential sampling ($\gamma = 0.5$), the frequency of "Security" increased from 19.6% ($v = 0$) to 36.1% ($v = 0.4$) and to 79.0% ($v = 0.9$); the frequency of "Development" also increased from 21.9% ($v = 0$) to 28.0% ($v = 0.4$), but it started decreasing because of "Security." The increase in the frequency of these two topics led to a decrease in the frequency of other topics. In weak sequential sampling ($\gamma = 0.25$), their frequencies changed in a similar way but by much smaller degrees; the only exception was the increase in the size of "Development" until $\upsilon = 0.8$.

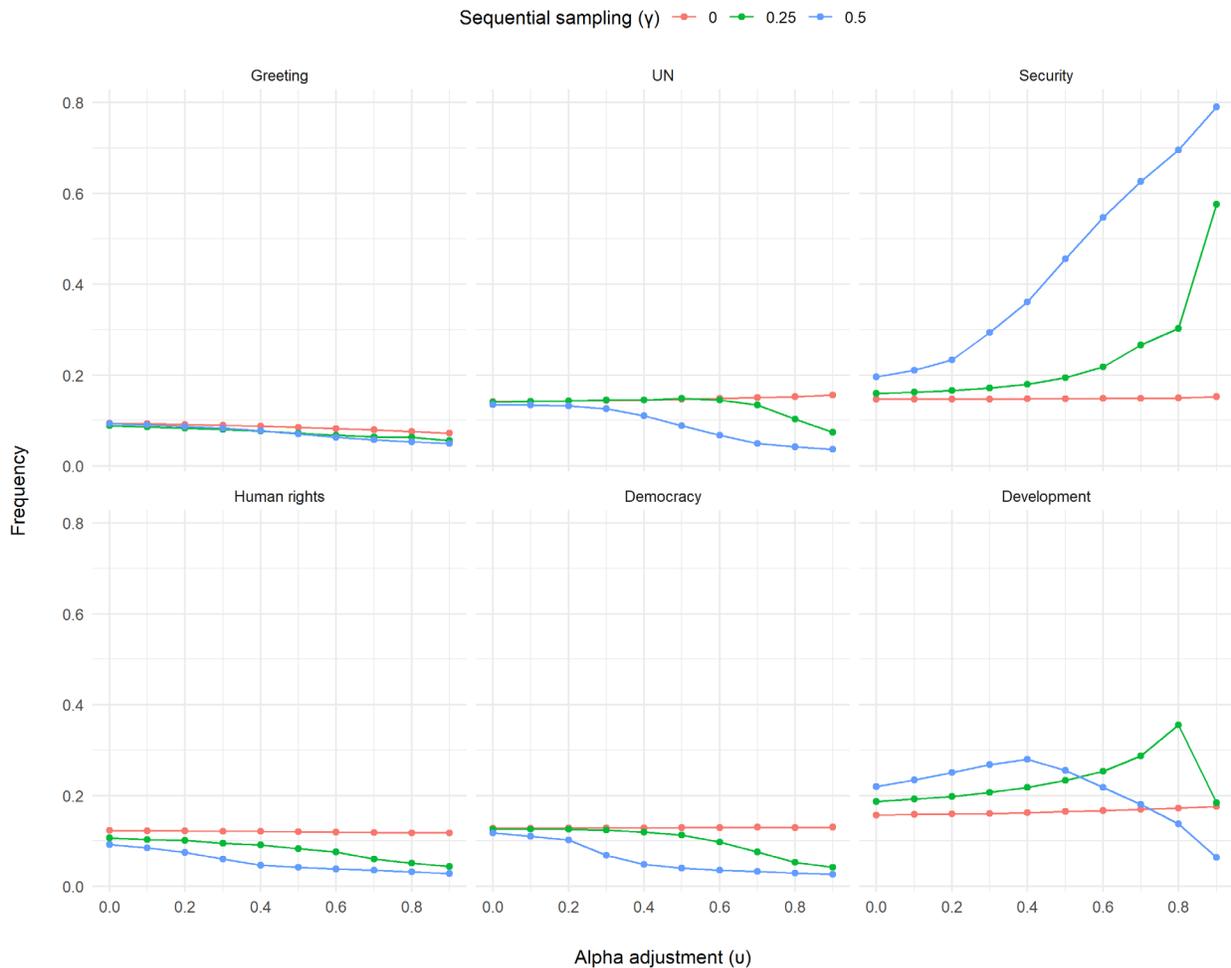

**Figure 8.** Topic frequency of DAA by different levels of Dirichlet prior adjustment. The *x*-axis is the maximum amount of adjustment made to $\alpha_k$; the *y*-axis is the frequency of topics in the corpus. The plotted lines represent the average values of the five models fitted under the same condition.

## Convergence Detection

The convergence detection greatly decreased the computational cost by interrupting the iterations early, but it did not adversely affect the F1 score or the perplexity score (Figure 9). The iteration was interrupted 200 times in 67% of the cases in the symmetric model and 74% of the cases in the asymmetric model. The number of iterations was sometimes greater but still less than or equal to 400 times in 92% of the cases in the former and 96% of the cases in the latter.

Despite much reduction in the number of iterations, the F1 and perplexity scores of the models reached the highest or lowest levels. The F1 score peaked when the number of iterations was around 200 times in both symmetric and asymmetric models, while the perplexity score fell sharply from 100 to 200 times and gradually from 200 to 500 times. Despite their significantly different F1 scores, their perplexity scores were roughly the same or slightly higher in the asymmetric models than in the symmetric models.

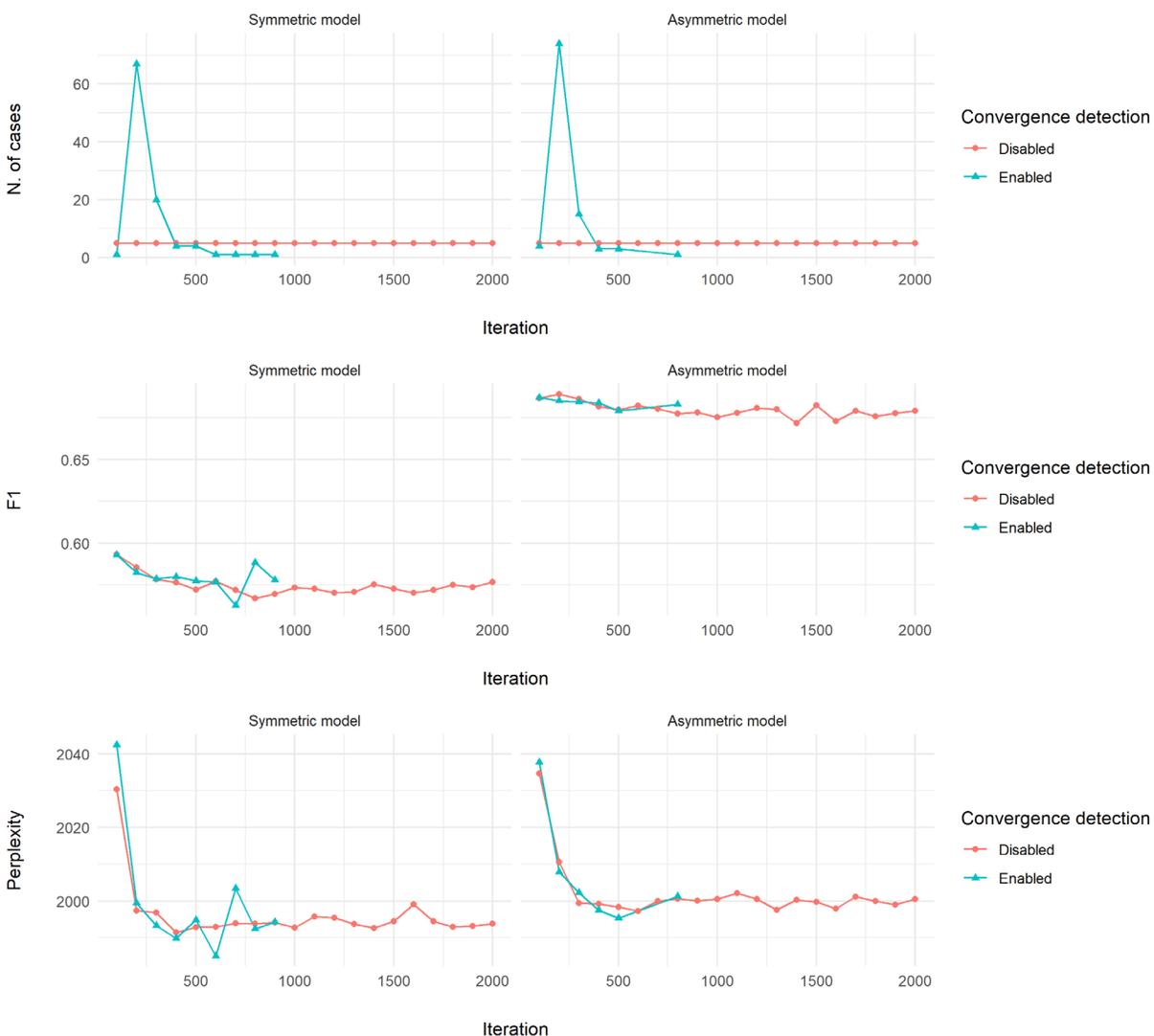

**Figure 9.** F1 and perplexity scores by the number of iterations. The *x*-axis is the number of iterations to fit the models; the *y*-axis is the number of cases (top), the F1 score (middle), or the perplexity score (bottom). The plotted lines represent the average values of the models fitted under the same condition. The other hyper-parameters are $v = 0$ and $\gamma = 0.25$ for the symmetric model and $v = 0.3$ and $\gamma = 0.25$ for the asymmetric model.

## Alternative Approaches

An alternative approach to topic classification of short documents is the clustering of documents represented in a lower dimensional space. Among various dimensional reduction techniques, I chose Doc2vec (Le and Mikolov 2014) here because it can be trained on the current corpus using an efficient distributed computing algorithm. I trained the Word2vec model (Mikolov et al. 2013) as the underlying model with different hyper-parameters (the length of vectors and the size of word window) and transformed the sentences into document vectors as the weighted average of the word vectors. I classified the document vectors into the pre-defined topics using k-means.[10]

The classification accuracy stayed largely constant regardless of the values of hyper-parameters (Figure 10). The overall F1 score was slightly higher than others when the vector length was 200 and the window size was 10. The F1 scores were comparable to the symmetric LDA models' only in "Greeting" and "Development". Despite the significantly higher computational costs, the long document vectors or the large windows did not improve the scores.

---

[10] See the Appendix for the detail of the clustering of document vectors.

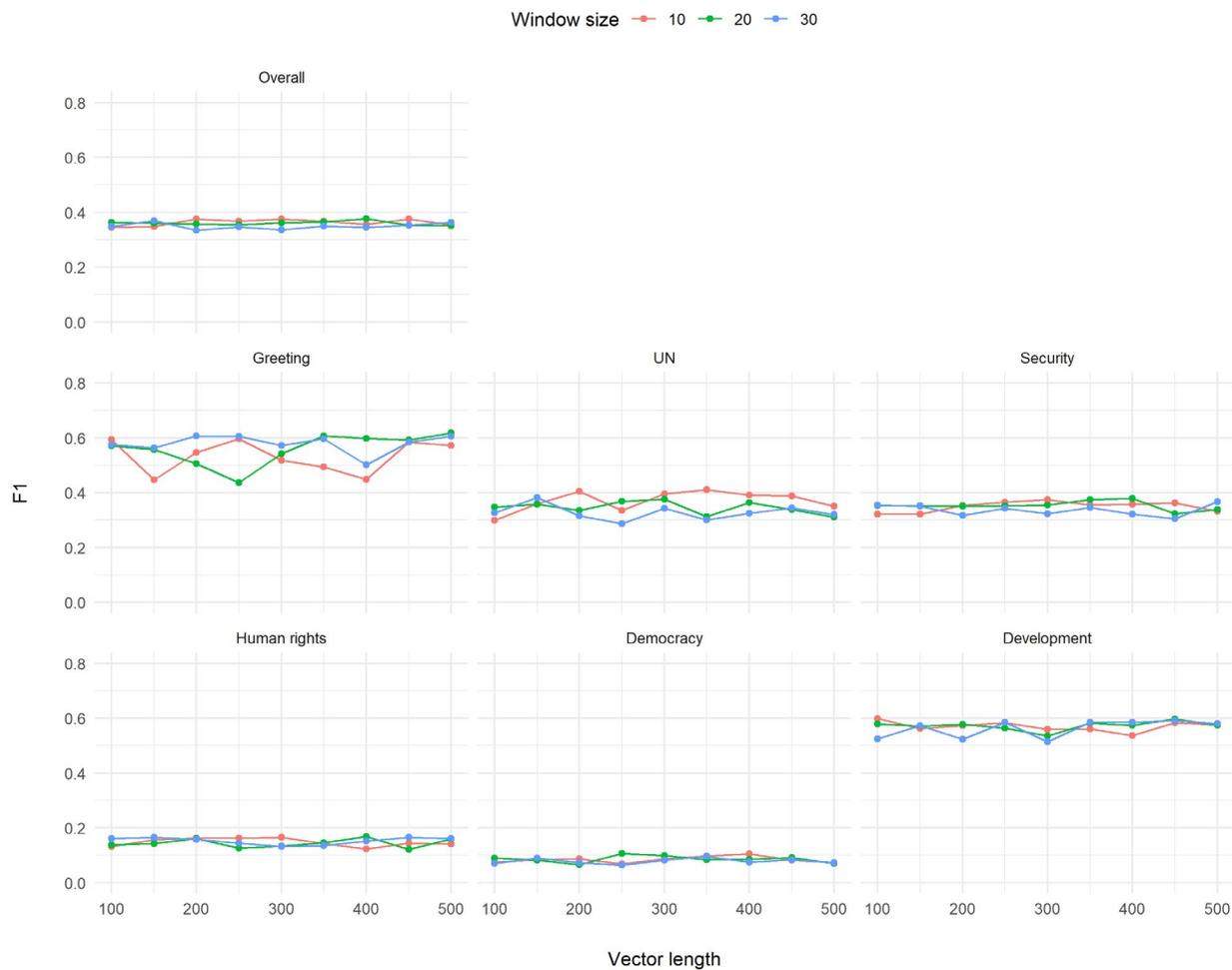

Figure 10: Classification accuracy of document vector-based clustering. The x-axis is the length of the document vectors; the y-axis is the micro-average F1 score. The plotted lines represent the average values of the five models fitted under the same condition.

## Example

I fitted DAA with symmetric or asymmetric Dirichlet priors to illustrate how they can lead to different conclusions in content analysis. Both models are fitted with the strong sequential sampling ($\gamma = 0.5$) and two unseeded topics for generic words; distributed computing and convergence detection are also enabled in both models. Moreover, the Dirichlet prior adjustment is enabled in the asymmetric model ($v = 0.3$) but disabled ($v = 0$) in the symmetric model to make it equivalent to distributed LDA.[11]

**Topic Terms**

The most frequent topic terms of the two models were very similar: words disagreed only in their order in "Greeting," "UN," "Security," and "Human rights" (Table 2). However, the symmetric model had "peace" in "Democracy" instead of "elections," which is more strongly related to democratic politics, and "terrorism" in "Other2" among generic words, which is clearly related to "Security." In both models, words with the same stems (e.g., "conflict" and "conflicts") were found in the same topic.

| Topic | Symmetric model | Asymmetric model |
|---|---|---|
| Greeting | mr, great, hope, wish, session, express, welcome, united nations, general assembly, also | mr, great, hope, wish, session, express, welcome, general assembly, united nations, also |
| UN | united nations, organization, security council, general assembly, session, reform, secretary-general, charter, conference, resolution | united nations, organization, security council, general assembly, session, reform, charter, secretary-general, conference, resolution |
| Security | peace, security, security council, weapons, nuclear, conflict, terrorism, international, war, conflicts | peace, security, international, security council, conflict, terrorism, weapons, nuclear, war, conflicts |
| Human rights | people, community, human rights, respect, women, responsibility, international, humanitarian, children, protection | people, community, human rights, respect, women, responsibility, humanitarian, children, protection, international |
| Democracy | government, republic, democratic, president, democracy, law, institutions, free, *peace*, freedom | government, republic, president, democratic, democracy, law, institutions, free, freedom, *elections* |
| Development | development, economic, cooperation, developing, social, countries, sustainable, poverty, trade, *progress* | development, economic, cooperation, countries, social, developing, sustainable, poverty, *international*, trade |

---

[11] By fitting the asymmetric model, the value of its Dirichlet prior, $\alpha_k$, were adjusted automatically from initial 0.5 to 0.449 in "Greeting", 0.489 in "UN", 0.714 in "Security", 0.405 in "Human rights", 0.420 in "Democracy", 0.681 in "Development", 0.421 in "Other1", and 0.420 in "Other2". The relative sizes of the values roughly correspond to the frequency of the topic in the corpus.

| Other1 | world, us, must, *new*, can, *international*, one, *global*, *countries*, *nations* | world, us, *people*, one, can, must, *today*, *many*, *country*, *united* |
| Other2 | *people*, world, *terrorism*, *country*, international, us, must, *united*, *states*, one | world, *new*, us, must, *can*, international, one, *global*, *nations*, *today* |

**Table 2.** Top 10 most frequent topic terms in the symmetric and asymmetric models. Italicized words highlight the difference between the models in each topic. "Other1" and "Other2" are unseeded topics.

## Topic Frequencies

The two models classified sentences in the corpus very differently (Figure 11). The annual frequencies of topics varied between 1,000 and 6,000 times in the symmetric model, but they ranged between 500 and 8,000 times in the asymmetric model. "Security" was nearly twice as frequent in the latter than in the former in the 1990s, while "Human rights" was roughly twice as frequent in the former than in the latter throughout the period. The frequencies of "Security" and "Development" were roughly equal between 1991 and 2005 in the symmetric model, but "Security" was significantly more frequent in the asymmetric model. Similarly, "Security" was less frequent than "Development" after 2005 in the former model, but they were roughly equal in the latter model.

The frequency of "Security" also correlates more strongly with key events such as the Kosovo war (1998–1999), the 9/11 attacks (2001), and the Arab Spring (2011–2012) in the asymmetric model compared with the symmetric model. The frequency of the topic was significantly higher during events in the former, but it was only marginally higher at the outset of the events in the latter. The results are more plausible in the asymmetric model than in the symmetric model.

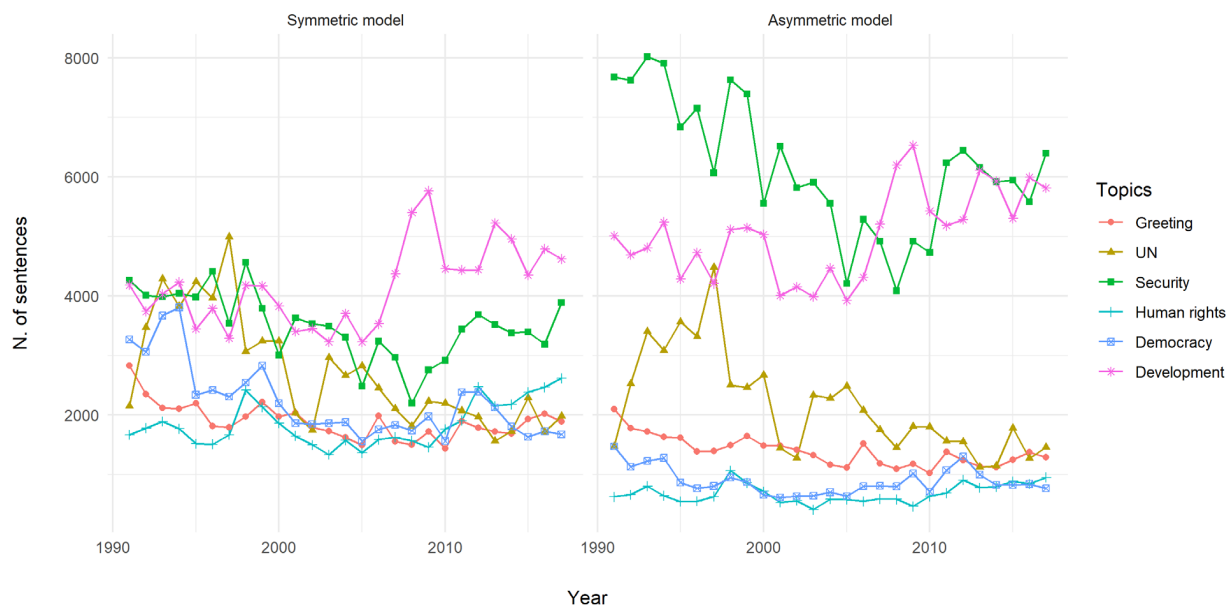

**Figure 11.** Topic frequency by year in the symmetric and asymmetric models. The *x*-axis is the year of speech; the *y*-axis is the number of sentences classified into topics. Dirichlet prior optimization is disabled in the symmetric model ($v = 0$) but enabled in the asymmetric model ($v = 0.3$).

## Discussion

The evaluation of the algorithms clearly showed that DAA can identify topics in the imbalanced corpus more quickly and accurately than LDA. DAA's distributed computing made inferences more than four times faster to identify 50 topics in the corpus when 8 processors were used; its Dirichlet prior optimization improved the F1 scores by 0.11 points overall and by more than 0.15 points in frequent topics. The use of Doc2vec could alleviate the data sparsity but did not contribute to higher F1 scores because clustering of document vectors is agnostic about the frequency of topics.[12]

---

[12] The overall frequency of topics is available through the frequency of seed words in the corpus, but this information is lost when sentences are transformed into document vectors.

DAA's convergence detection reduced the number of iterations from 2000 to 200 times in more than 90% of the cases while achieving the highest level of F1 scores. The example also demonstrated that the results of content analysis are vastly different between DAA and LDA. These differences were caused by the tendency of the symmetric model to under or overestimate the frequencies of topics, reaching nearly 50% less in the most frequent topics and 100% more in the least frequent topics compared with the asymmetric model's estimates.

The advantage of integrating multiple algorithms, as in DAA, was seen in the interactions between algorithms. Distributed computing had a limited impact when the number of topics was small, but convergence detection reduced the computational cost by 10-fold even in such cases. The Dirichlet prior adjustment alone did not have much of an impact, but it significantly increased the F1 scores when sequential sampling was enabled. However, the classification accuracy of DAA fell sharply because strong Dirichlet prior adjustment ($0.4 \leq \upsilon$) and sequential sampling led to the rapid increase in the frequency of "Security." This suggests that these algorithms help the model to identify imbalanced topics more accurately, but interactions between them can cause an oversampling of frequent topics. To avoid this, users should limit the adjustment between $0.1 \leq \upsilon \leq 0.3$ when strong sequential sampling is used.[13]

It appeared difficult to compare fitted LDA and DAA models in terms of the goodness-of-fit because their perplexity scores were roughly the same, but the noticeable differences between them in top topic terms suggest it is possible. The differences between the symmetric and asymmetric models were clearer in infrequent topics (i.e., "Democracy" and "Other2") because symmetric Dirichlet priors caused misassignment of infrequent topics to frequent words, which then received high probability scores in these topics. To assess the quality of models fitted on an imbalanced corpus, users should inspect the top topic terms focus on infrequent topics. If words that relate to more frequent topics are found, the model needs greater adjustment in terms of the Dirichlet priors (i.e., a greater value of $v$).

---

[13] The initial value of the Dirichlet prior in this study is set at $\alpha = 0.5$ because the number of topics was only eight, but it must be smaller when the number of topics is larger.

Finally, the ability of DAA to identify highly specific topics in large corpora enhances the reliability of topic analysis by reducing arbitrary operations before and after fitting models. It has been common to perform aggressive feature engineering such as stemming, but it is no longer necessary because DAA can assign the same topic to different forms of a word (i.e., inflections). It has also been common to group topics either manually or automatically to create more meaningful clusters, but it is no longer justified because fragmentation of topics can be avoided using DAA.

**Conclusions**

I have developed DAA by integrating various algorithms that enhance LDA, implemented it in an open-source software package, and demonstrated its ability to identify topics in a large corpus of short documents. It may not outperform topic models that utilize pre-trained word vectors, but its improved performance suggests that LDA can be further developed and used in social science research. By using LDA-based models, social scientists can ensure their analyses are transparent and independent from third-party tools.

Social scientists must also be aware of the complexity of statistical modeling of topics in an imbalanced corpus to improve the quality of their analyses: the Gibbs sampler must be informed on the overall frequencies for topics when they are unequal. This means that users of LDA must pay more attention to the Dirichlet priors and optimize their values to perform content analysis more accurately. If they cannot accomplish this manually, it should be done automatically using DAA instead.

This study broadly contributes to social sciences by creating new algorithms for convergence detection and Dirichlet prior adjustment. It integrates them with other algorithms that enhance the speed and accuracy of LDA. However, it remains challenging to find the optimal level of sequential sampling and Dirichlet prior adjustment based on the perplexity scores. In future research, a new measurement that helps users choose the optimal value of the hyper-parameters needs to be developed.

**Data availability statement**

The data and scripts used in this study is available at https://doi.org/10.7910/DVN/2S1XKZ.

**Declaration of conflicting interest**

The author declared no potential conflicts of interest with respect to the research, authorship, and/or publication of this article.

## Appendix

### Clustering of document vectors

Transformation of sentences into document vectors is advantageous in clustering because they are dense representation of textual data. Although many scholars use pre-trained model such as BERT for this purpose, I trained Word2vec (Mikolov et al. 2013) to evaluate the algorithm's ability to learn topics from the local corpus. Word2vec is trained with the continuous bag of words (CBOW) algorithm because the algorithm iteratively optimizes word vectors based on the document vectors for words within the window. I created seed vectors as the average of word vectors for the seed words to pre-define (Table 1). These seed vectors are used as the initial centroids in k-mean clustering (Forgy 1965) along with two random centroids for unseeded topics. K-means are fitted through 100 iterations.

To identify words associated with topics, the frequency of words is aggregated and weighed by the TF-IDF treating the clusters as documents (i.e., cluster TF-IIDF); then words are sorted in the descending order by their weighted frequency in each cluster (Table 3). The topic words in "UN", "Security" and "Development" are highly relevant but they are all proper names in "Greeting", specifically about the Arab-Israeli conflict in "Human rights" and strongly related to security in "Democracy".  Topic words in "Other1" are generic but those in "Other2" are about development.

| Topic | Words |
| --- | --- |
| Greeting | opertti, essy, ganev, thomson, amara, ping, didier, diogo, freitas, amaral, hennadiy, nassir, seung-soo, theo-ben, insanally, harri, treki, al-nasser, udovenko, abdussalam |
| UN | representativeness, composition, standby, veto, system-wide, viii, commissioner, streamlining, stand-by, revitalize, avoid, norms, counter, categories, peace-enforcement, enlarging, coordinating, democratized, deliberative, duplication |
| Security | maimed, civilians, coup, condolences, kill, massacres, injured, serbian, killed, illegal, rape, kashmiri, cleansing, golan, mosque, religion, settlers, detention, repression, rockets |
| Human rights | non-interference, norms, two-state, arab-israeli, al-quds, al-sharif, tunb, non-intervention, religion, jerusalem, neighbourliness, equity, violated, golan, reaffirms, israelis, violates, avoid, karabakh, harmonious |
| Democracy | nuclear-test-ban, chemical, rarotonga, tunb, weapon-free, cut-off, ministerial, referendum, test-ban, six-party, scottish, prohibition, g-8, tlatelolco, ceasefire, plo, deposit, envoy, golan, indefinite |
| Development | post-2015, emissions, sdgs, sids, employment, low-carbon, fiscal, oda, renewable, liberalization, competitiveness, technological, vocational, resource, almaty, conservation, monterrey, holistic, forestry, equity |
| Other1 | poet, divine, philosopher, bless, avoid, ye, interdependent, love, clash, beauty, colour, creator, prophet, prophets, taught, naive, religion, technological, allah, bipolar |

| Other2 | prices, emissions, rates, fossil, gdp, income, inflation, mortality, agricultural, trillion, tourism, products, estimates, flows, carbon, illegal, estimated, earnings, commodity, drinking |

**Table 3**: Topic words in k-mean clustering. The length of the document vector and the size of the window are set to 200 and 30, respectively, for the Doc2vec model.